\documentclass[aps,twocolumn,superscriptaddress,floatfix,prl,longbibliography]{revtex4-2}
\usepackage{float}
\usepackage{mathrsfs}
\usepackage{graphicx}
\usepackage{amsmath,amssymb}
\usepackage{multirow}
\usepackage{subfigure}
\usepackage{braket}
\usepackage{bm}
\usepackage{color}

\begin{document}
\title{Superconductivity of bipolarons from quadratic electron-phonon interaction}

\author{Zhongjin Zhang}
\affiliation{Department of Physics, University of Massachusetts, Amherst, MA 01003, USA}

\author{Anatoly Kuklov}
\affiliation{Department of Physics \& Astronomy, College of Staten Island and the Graduate Center of
CUNY, Staten Island, NY 10314}

\author{Nikolay Prokof'ev}
\affiliation{Department of Physics, University of Massachusetts, Amherst, MA 01003, USA}

\author{Boris Svistunov}
\affiliation{Department of Physics, University of Massachusetts, Amherst, MA 01003, USA}

\begin{abstract}
In systems with linear electron-phonon interaction (EPI), bound states of polarons, or bipolarons, form by gaining energy from the lattice deformation. The quadratic EPI case is fundamentally different: bipolarons form because electrons lose less energy when the total charge density is ``compacted." As the coupling constant is increased,
the bipolarons first appear as extended (but finite radius) 
soliton-type states. They subsequently decrease in radius until
their size reaches the inter-atomic scale. We present the first numerically exact solution of the bipolaron problem from quadratic EPI in the presence of both on-site Hubbard and long-range Coulomb repulsion, and compute estimates of the largest superconducting transition temperature within the bipolaron mechanism. We find that $T_c/\Omega$
ratios, where $\Omega$ is the optical phonon frequency, 
can be several times larger than what one may expect from the linear EPI provided the phonon frequency is increased by orders of magnitude on occupied sites. Electron-electron repulsion can be tolerated at the expense of stronger EPI and the most detrimental effect comes from the Coulomb potential because it easily eliminates extended soliton states.  
\end{abstract}

\maketitle


When electrons couple to lattice vibrations their low-energy 
properties such as the dispersion, $\epsilon ({\mathbf k})$, effective mass, $m_*$, and residue, $Z$, get renormalized; the 
resulting quasiparticle is called a polaron \cite{Landau1}. 
In the strong coupling limit, the electron-phonon interaction (EPI) 
can also cause two polarons to form a bound state of size $R$, or a bipolaron. At finite but low enough density, when bipolarons can be treated as point bosons with repulsive effective interaction, the transition temperature to the superconducting state can be accurately estimated from (see Ref.~\cite{PhysRevLett.130.236001})
\begin{equation}
T_c(n_b) \approx 3.31 n_{b}^{2/3}/m_* \, .  
\label{Tc3Dn}     
\end{equation}
The maximum value of $T_c$ within the bipolaron mechanism follows from the 
``marginal overlap" condition on the bipolaron density $n_b$, 
namely $4\pi R^3 n_b/3 =1$. On a lattice, this condition makes sense only 
if $R >a$ where $a$ is the lattice constant. When two electrons form a bound state localized on a single site, one formally has $R \ll a$ but the density is still 
limited because bipolarons cannot occupy the same site. To reflect this fact, we 
introduce effective radius squared, $R_*^2 = \max{(R^2,a^2)}$, and use it instead of $R^2$ in the no-overlap condition. This leads to the following estimate for the largest transition temperature \cite{PhysRevLett.130.236001}
\begin{equation}
T_c \sim  \frac{1.3}{m_* R_*^2} \, ,
\label{Tc3DR} 
\end{equation}
in terms of single bipolaron parameters. This estimate clearly shows that 
high values of $T_c$ require light and compact bipolarons. 

Nearly all work on bipolarons in the past was limited to linear
(in terms of atomic displacements) EPI when bound states form  
by gaining energy from large lattice distortion. In the most
relevant for materials adiabatic limit $\Omega /W \ll 1$, 
where $\Omega$ is the characteristic phonon frequency and $W$ is the
bare electron bandwidth, this interaction 
mechanism leads either to exponentially heavy (interatomic size) 
bipolarons or large-radius shallow states with sharp 
crossover between the two \cite{Chakraverty,Chao2}. Systems 
with phonon-assisted hopping EPI have larger $T_c/\Omega$ ratios 
than systems with density-displacement EPI, but in all known cases, this
ratio remains much smaller than unity \cite{Chao2,Chao3} in the adiabatic limit. 

This outcome motivates one to consider alternative mechanisms of EPI, 
and one promising example is the quadratic density-displacement coupling
\begin{equation}
H_{\rm ep} \, =\,  g_2 \sum_{i} n_{i} \frac{M\Omega^2 x_i^2}{2} \, \equiv \,
g_2\frac{\Omega}{4}\sum_{i} n_{i} \left( b_i^{\,}+b_i^\dagger \right)^2 , 
\label{Hint}
\end{equation}
which results in light compact polarons \cite{X2polaron,X2soliton}.
Here $b_i$ is the bosonic annihilation operator on site $i$, $n_{i}$ is the 
electron occupation number, $n_i=\sum_{\sigma }n_{i \sigma}$, 
and $g_2>0$ is the dimensionless coupling constant. 
The effects of weak quadratic coupling to phonons in the Fermi liquid regime were considered numerous times in the past \cite{soft1,Entin,Riseborough,Entin2,Hizh,Heid,Matthew,Mahan,detMC2,soft4,soft3,Cava,soft2,Andy1,Andy2,x2materials1,x2materials2,x2materials3,x2materials4}. The soliton-type polaron \cite{Kuklov,Gogolin, Gogolin2,Mona00,X2polaron,X2soliton}  and bipolaron \cite{Kuklov2} formation within the variational approach in continuum was addressed too, and it has been realized that the effective mass is not exponentially suppressed at strong coupling. However, 
so far the entire strong coupling regime for bipolarons and 
optimal $T_c$ values have not been solved. Recently, a special case---the atomic limit (AL) in the absence of interelectron repulsion was investigated in Ref.~\cite{PhysRevLett.132.226001}. 
[The AL regime corresponds to large values of $g_2$ when both electrons reside on one and the same lattice site.]
The most important result of this study is that, similarly to single polarons \cite{X2polaron,X2soliton}, 
the bipolaron mass in this extreme limit remains moderately enhanced and 
weakly dependent on the coupling constant. 
However, the physics of extended soliton states,
which form well before the AL is reached, suggests that optimal values of $T_c$ are expected for soliton states before they collapse to the AL because 
compact solitons are light and their masses are nearly independent of $g_2$ (or even decrease with increasing coupling \cite{X2soliton}). 

The non-interacting part of the  Hamiltonian considered in this work describes 
electrons with nearest neighbor hopping on the simple cubic lattice 
and dispersionless optical phonons (we count energy from the ground state of 
non-interacting harmonic oscillators):
\begin{equation}
H_0 =  -t \sum_{\braket{ij},\sigma}c_{j\sigma}^\dagger c_{i\sigma}^{\,} +\Omega \sum_i b_i^\dagger b_i^{\,} .
\label{H0}
\end{equation}
Here $c^\dagger_{i\sigma}$ ($c^{\,}_{i\sigma}$) create (annihilate)  electron with spin $\sigma$ on site $i$. The value of $t$ is chosen as the unit of energy, and the lattice constant $a$ as the unit of length.
Apart from EPI (\ref{Hint}) the 
interacting Hamiltonian also includes direct electron-electron repulsion characterized by the on-site Hubbard term $U$ and long-range Coulomb potential $V(r>0)=V (a/r)$:     
\begin{equation}
    H_{\rm ee} \, = \, U\sum_in_{i\uparrow}\, n_{i\downarrow}\,  +\,  
                 \sum_{i\ne j } \, V(| r_i-r_j |) \, n_{i\uparrow} n_{j\downarrow} \, .
\label{Hee}
\end{equation}

According to Hamiltonian (\ref{Hint}), the phonon frequency on sites with the occupation number $n$ is changed to
\begin{equation}
\tilde{\Omega}_n=\Omega r_n\;, \qquad r_n =\sqrt{1+g_2 n }\,. 
\label{On}
\end{equation}
If $g_2$ is negative (the model is stable only at $g_2\ge -1/2$) 
the energy gain due to change in the harmonic oscillator zero-point energy 
on doubly occupied site is larger than that on two distant 
singly occupied sites
\begin{equation}
U_{\rm eff} = \frac{\Omega }{2} [ r_2-1 - 2(r_1 - 1) ] = 
\frac{\Omega }{2} [r_2+1-2r_1] \leq 0   \,.
\label{Ueff}
\end{equation}
In the adiabatic limit, this weak attraction (even in the absence of Hubbard and Coulomb repulsion) cannot bind electrons in three dimensions. 
However, $U_{\rm eff}$ is negative also for positive $g_2$ because energy increase with electron density is sub-linear. This is the prime reason why polarons at strong coupling form extended
self-trapped soliton states \cite{Kuklov,Kuklov2,Gogolin, Gogolin2,X2soliton}.
It is then expected that bipolarons first appear as extended finite-radius solitons; 
the radius decreasing with $g_2$ until the AL is reached at approximately
$U +U_{\rm eff} < -W/2) $ (in the simple cubic lattice $W=12t$), 
i.e. at 
\begin{equation}
g_{\rm AL} \, \sim \, 3 \left( \frac{W+2U}{\Omega} \right)^2 . 
\label{gAL}
\end{equation}
 We encounter large values of $g_2$ in this work, which correspond to  a situation when the phonon frequency softens as $\Omega \to 0$ close to the structural phase transition in the absence of electrons, while at a finite electron numbers $n=1,2$ its value $\sqrt{ng_2}\Omega$ undergoes no softening. This naturally implies that $g_2$ can reach values $ \gg 1$ and ultimately $\to \infty$.

\begin{figure}[htbp]
			\begin{center}
               \includegraphics[width=0.99 \columnwidth]{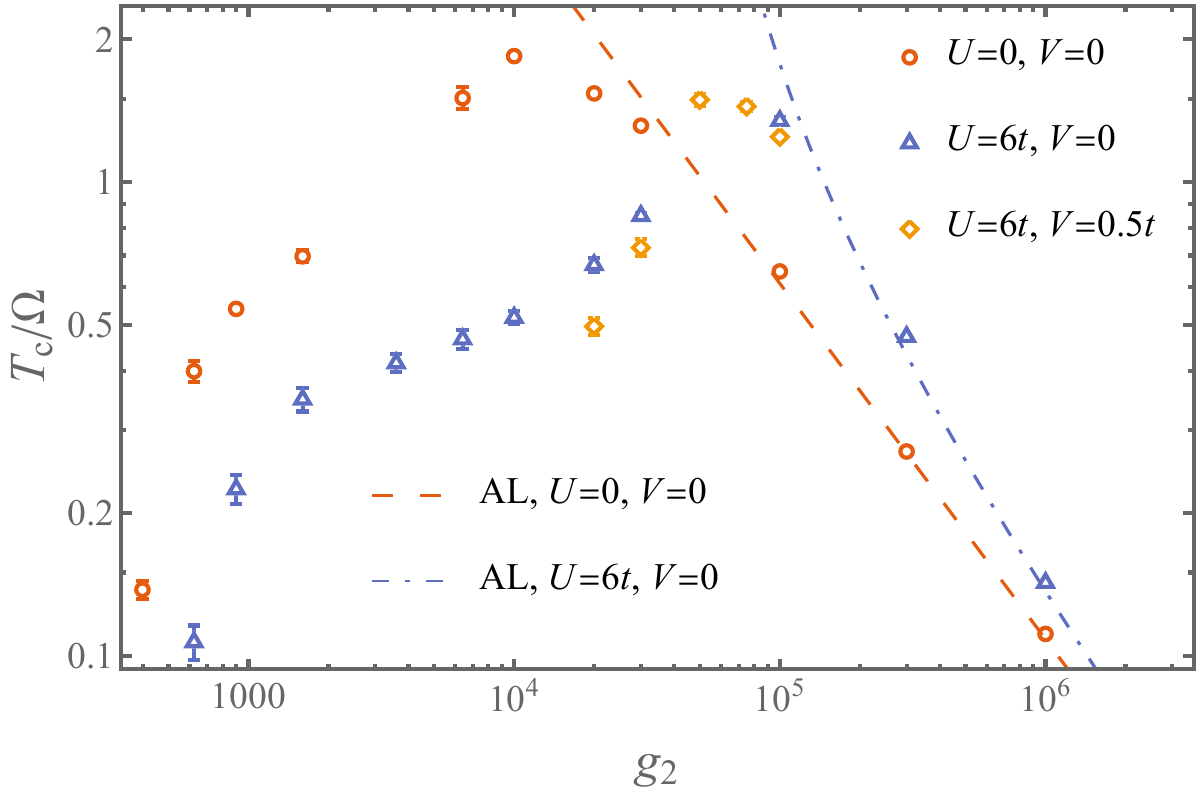}
                \end{center}
\caption{ The critical temperature to phonon frequency ratio as a function of the 
coupling strength $g_2$ with and without onsite and/or Coulomb repulsion.
Dashed lines are asymptotic predictions for the atomic limit. Bound states 
disappear for $g_2$ smaller than the leftmost point for each case, see Fig.~\ref{dEvsg2}. }
\label{Tcvsg2}
\end{figure}

In this work, we first look into the atomic limit $g_2 > g_{\rm AL}$,  add the
effects of electron-electron interaction to the theoretical description, and 
use it as a benchmark for numerical simulations of the effective mass, the bound state size, and $T_c$ estimates in this limit. We perform numeric simulations of the system Hamiltonian in the basis of atomic coordinates using the $X$-representation Monte Carlo technique (XMC)---a numerically exact sign-free approach \cite{Xrep,X2polaron,X2soliton}. We obtain accurate results for bipolarons in the entire strong coupling regime for three representative cases (with and without on-site and/or Coulomb interaction) in the 
realistic adiabatic regime $\Omega /t =0.1$ or $\Omega /W = 1/120$. 
We find that the optimal values of $T_c /\Omega $ ratio 
exceed the largest known results based on the linear EPI by several times,
and are found on approach to the AL as a result of competition between the  
shrinking soliton size and increasing mass. The main result is presented in 
Fig.~\ref{Tcvsg2}.


\textit{Method}. 
The XMC method is based on the lattice path integral for the 
electrons and coordinate representation for harmonic oscillators \cite{Xrep,X2polaron,X2soliton}.  
On the one hand, lattice path-integral allows us to deal with arbitrary 
interaction potential between the electron because $H_{\rm ee}$ is 
diagonal in this basis and is kept in the exponent. On the other hand,
in the coordinate representation of harmonic oscillators, all EPI effects 
are absorbed into the harmonic oscillator propagator with the electron occupation number dependent frequency (\ref{On}). All relevant bipolaron properties are extracted from
the electron pair propagator in imaginary time  
\begin{equation}
    G(\tau,\bm r_1,\bm r_2)=\braket{{\rm vac}|c_{\bm r_1\uparrow}c_{\bm r_2\downarrow}{\rm e}^{-\tau H}c^\dagger_{\bm 0\downarrow}c^\dagger_{\bm 0\uparrow}|{\rm vac}}.
\label{G2}    
\end{equation} 
simulated using Monte Carlo sampling methods. From asymptotic exponential decay of 
$G_p(\tau)$, which is the Fourier transform of the center-of-mass propagator $G(\tau,\bm R) =\sum_{\bm r_1}\sum_{\bm r_2} \delta_{\bm R,(\bm r_1+\bm r_2)/2}G(\tau,\bm r_1,\bm r_2)$, as a function of imaginary time, we extract the bipolaron energies in different momentum sectors: $G_p(\tau \to \infty) \sim Z_p {\rm e}^{-E_p\tau}$. 
This gives us direct access to the ground-state energy, $E = E_{p = 0}$; the effective mass, $1/m_* = {\rm d}^2 E_p/ {\rm d} p^2$, is extracted by fitting the low-momentum data.
Other observables in the ground state are measured in the middle 
of the path \cite{Chao2} using asymptotically exact in the $\tau \to \infty$ 
limit relation
\begin{equation}
 \braket{G|\hat{O}|G}= \frac{\braket{{\rm vac}|c_{\bm r_1\uparrow}c_{\bm r_2\downarrow}{\rm e}^{-\tau H/2}\hat{O}{\rm e}^{-\tau H/2}c^\dagger_{\bm 0\downarrow}c^\dagger_{\bm 0\uparrow}|{\rm vac}}}{ \braket{{\rm vac}|c_{\bm r_1\uparrow}c_{\bm r_2\downarrow}{\rm e}^{-\tau H}c^\dagger_{\bm 0\downarrow}c^\dagger_{\bm 0\uparrow}|{\rm vac}}} \,.
 \label{Operator}
\end{equation}
For the operators which are diagonal in the chosen representation, such as electron positions, 
the computation of (\ref{Operator}) is simple: all one needs to do is to collect statistics of the matrix elements $O(\bm r_1, \bm r_2)$ of the operator $\hat{O}$ in the coordinate representation. The only minor setup requirement is that 
the overlap between the initial and ground states, $\braket{G|c^\dagger_{\bm 0\downarrow}c^\dagger_{\bm 0\uparrow}|{\rm vac}}$, is not too small to produce 
reliable XMC statistics for propagator (\ref{G2}) at large $\tau$.

All simulations in this work were performed in the realistic adiabatic limit with
$\Omega /t = 0.1$. We considered three representative cases: $(U,V) = (0,0)$, $ (6t,0)$, and  $ (6t,0.5t)$; and performed broad scans in terms of the coupling constant $g_2$.
Apart from the bipolaron effective masses and ground state energies
(to determine binding energies, $\Delta E = E-2E_p$, 
we had to simulate single-polaron ground state energies $E_p$ as well), 
we have computed probabilities, $P(\bm r)$, of finding two electrons at a given 
distance $\bm r$ from each other, which was subsequently used to obtain
the mean square bipolaron radius
\begin{equation}
R^2 = \braket{ r^2} = \sum_{\bm r} r^2 P(\bm r)\,.
\label{R2def}    
\end{equation}


\textit{Atomic Limit}. 
We first look at the limiting case $g_2 \gg g_{\rm AL}$ considered in Ref.~\cite{PhysRevLett.132.226001} when the binding state is localized 
on a single site and generalize the theory to include effects of on-site and Coulomb repulsion. Even for modest adiabatic ratio $\Omega /t =0.1$ considered in this
work, the AL corresponds to very large values of the coupling constant $g_2$ between 
$10^4$ and  $10^5$ depending on $U$ [see Eq.~(\ref{gAL}) and the comparison with the numerically exact solutions below]. 
When two electrons occupy the same site their energy is $ E_{\ell 1} = (\tilde{\Omega}_2-\Omega)/2+U = \Omega (r_2-1)/2+U$. If electrons occupy
different sites a distance $r$ apart, their energy is 
$ E_{\ell 2}(r) = \tilde{\Omega}_1-\Omega + V(r) = \Omega (r_1-1)+V(r)$.  
In the AL, the leading binding energy term is then given by the difference between the energies of two localized solutions with $r \to \infty$
\begin{equation}
\Delta E_{\ell } =  E_{\ell 1} - E_{\ell 2}(r\to \infty )
= U  + \frac{\Omega}{2} [r_2+1-2r_1] \,.
\end{equation}
 
The motion and zero-point fluctuations of tightly bound electrons 
are through second-order in $t$ processes when in virtual states 
one finds two electrons on the nearest-neighbor sites
with excited oscillator modes. The final result for the effective bipolaron 
hopping amplitude, $t_{\rm eff}$, and zero-point fluctuation energy, $\delta E$, is a straightforward generalization of calculation 
done in Ref.~\cite{PhysRevLett.132.226001}
\begin{equation}
t_{\rm eff}  =  \frac{2\eta t^2}{\Delta}  
\int_0^{+\infty} \frac{{\rm e}^{-z}\, {\rm d} z }
{1-\gamma_0\gamma_2{\rm e}^{-2\xi z}} \,,
\label{teff}
\end{equation}
\begin{equation}
\delta E = - \frac{12 \eta t^2}{\Delta}  
\int_0^{+\infty} \frac{ {\rm e}^{-z} \, {\rm d} z}
{\sqrt{1-\gamma_0^2{\rm e}^{-2\xi z}}
 \sqrt{1-\gamma_2^2{\rm e}^{-2\xi z}}  } \,,
\label{Ezp}
\end{equation}
where
\[
\Delta = \Omega [2r_1-r_2 -1]/2 -U+V \,,
\]
\[
\eta = \frac{4\sqrt{r_2} r_1}{(r_1+1)(r_2+r_1)} \,, \qquad \gamma_m  = \frac{r_1-r_m}{r_1+r_m}\,,
\]
and $\xi = r_1 \Omega /\Delta $ (here we correct the typing mistake 
in Ref.~\cite{PhysRevLett.132.226001}). 
With hopping corrections included, the bipolaron binding energy 
is given by 
\begin{equation}
\delta E =\Delta E_{\ell} +\delta E - 6t_{\rm eff} + \frac{24\sqrt{r_1}t}{1+r_1} \;,    
\label{EAL}
\end{equation}
where the last term is the kinetic energy gain for two mobile polarons. 
The effective mass renormalization is simply $m_*/m = t/t_{\rm eff}$. 

Finally, for direct comparison with numerical simulations we present 
the AL prediction for the bound state radius defined by Eq.~(\ref{R2def}): 
\begin{equation}
\begin{aligned}
 \left(\frac{R}{a}\right)^2  &=&  \frac{12 \eta t^2}{\Delta^2} 
 \int_0^{+\infty} {z\rm e}^{-z} {\rm d} z \left[
 \frac{1}{1-\gamma_0\gamma_2{\rm e}^{-2\xi z}}  \right. \\
 &+&  \left. 
\frac{1}{\sqrt{1-\gamma_0^2{\rm e}^{-2\xi z}}\sqrt{1-\gamma_2^2{\rm e}^{-2\xi z}}} 
\right] . \quad
\end{aligned}
\label{R2AL}  
\end{equation}
It has no effect on $T_c$ estimate based on $R_*^2 = \max(R^2,a^2)$ because in this
limit, $R\ll a$, and, thus, the transition temperature is decreasing with increasing  $g_2$ as $1/m_*$. In the asymptotic $g_2 \to \infty$ limit, we have 
$E \simeq E_{\ell 1} - 180 g_2^{-3/4}t^2/\Omega$,  $m_*/m \simeq 0.076 g_2^{3/4}\Omega/t$, 
and  $(R/a)^2 \simeq 550 g_2^{-5/4} t^2/\Omega^2 $.

\begin{figure}[htbp]
			\begin{center}
			\includegraphics[width=0.99 \columnwidth]{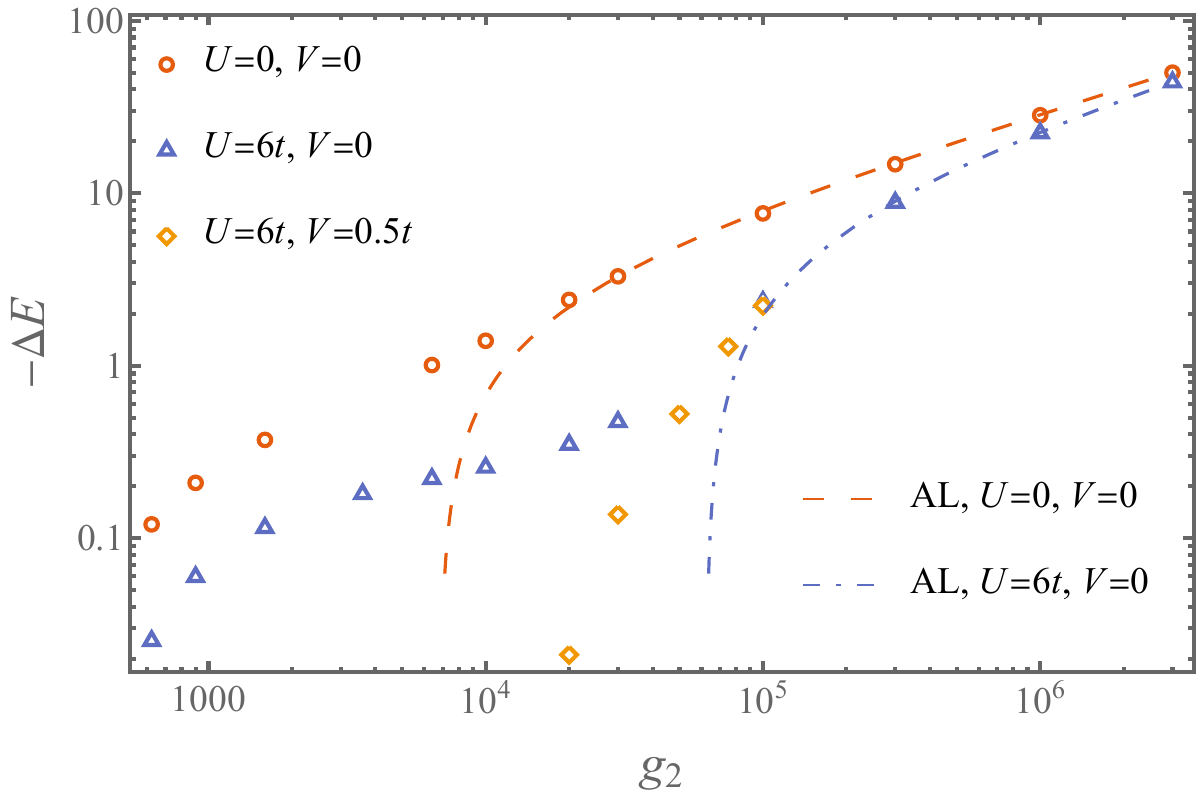}
                \end{center}
\caption{Bipolaron binding energies as functions of $g_2$ (data points).  
Dashed and dashed-dotted lines are the results of AL, Eq.~(\ref{EAL}). Statistical 
error bars are much smaller than symbol sizes.}
\label{dEvsg2}
\end{figure}
\begin{figure}[htbp]
			\begin{center}
               \includegraphics[width=0.99 \columnwidth]{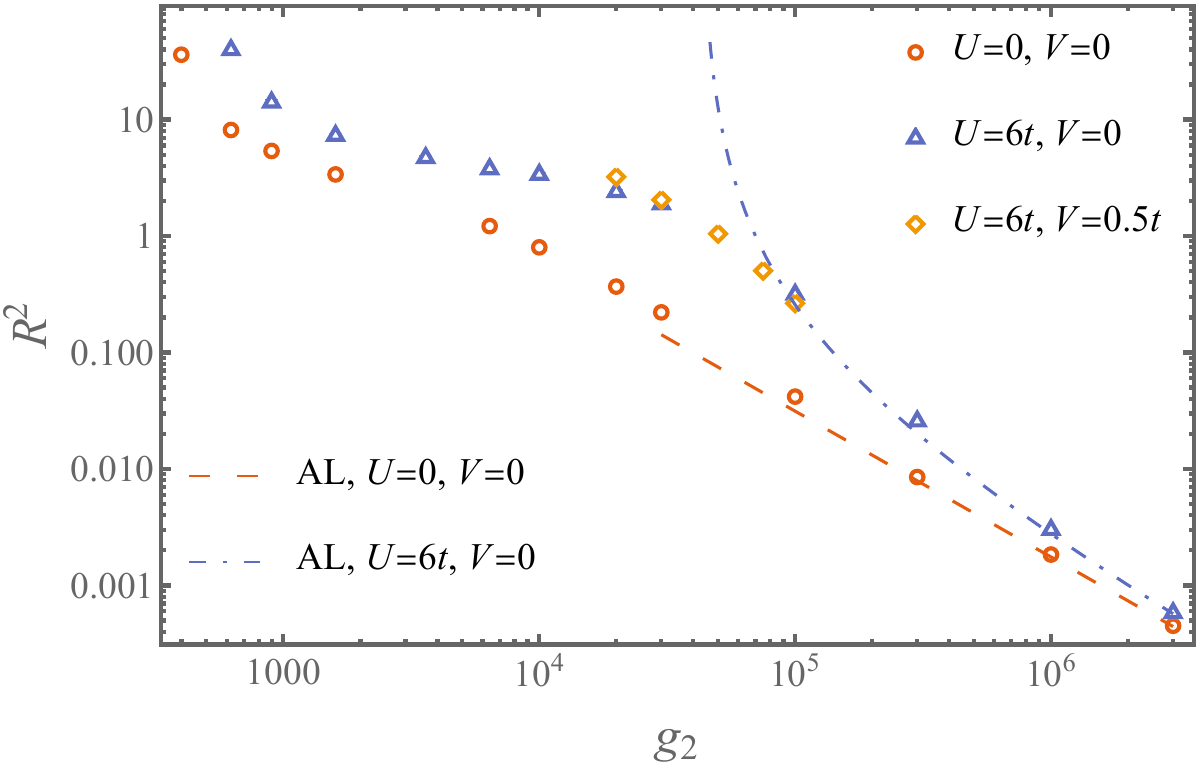}
                \end{center}
\caption{ Bipolaron radii as functions of $g_2$ (data points). 
Dashed and dashed-dotted lines are the AL results, Eq.~(\ref{R2AL}).
Statistical error bars are much smaller than symbol sizes.}
\label{Rvsg2}
\end{figure}
\begin{figure}[htbp]
			\begin{center}
               \includegraphics[width=0.99 \columnwidth]{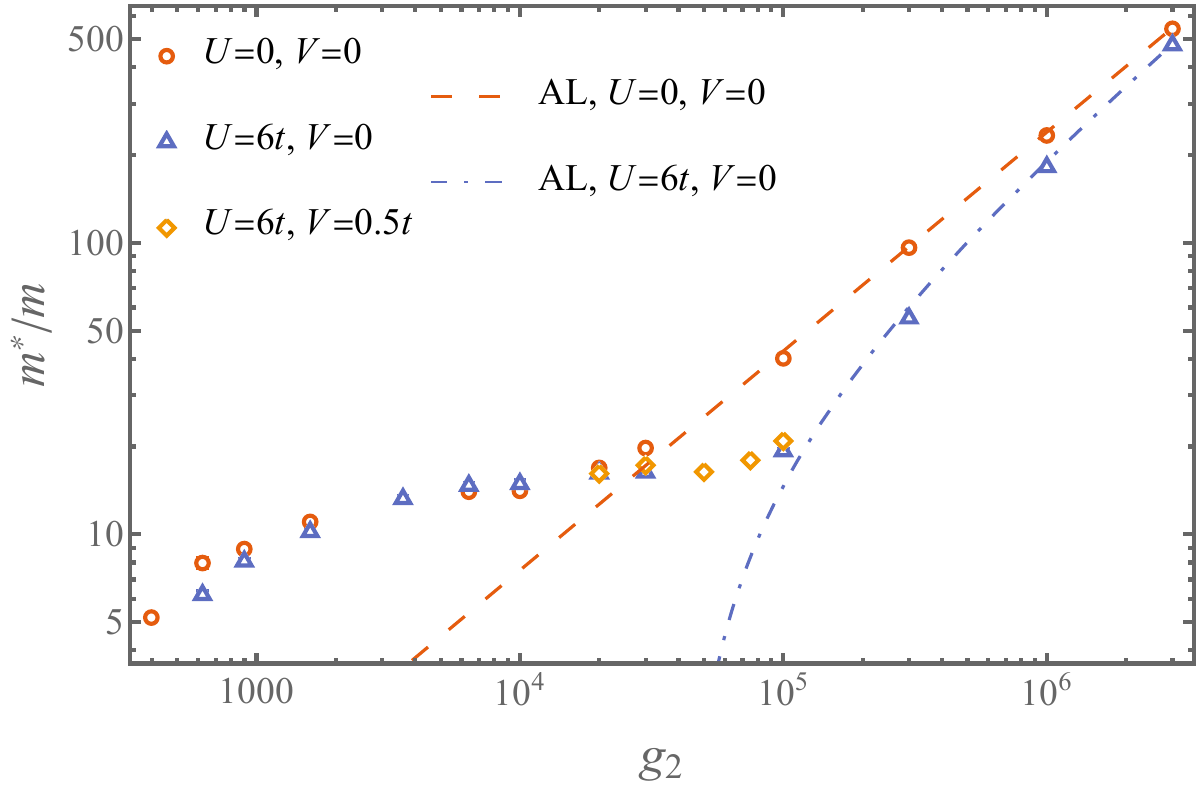}
                \end{center}
\caption{Effective masses as functions of $g_2$ (data points). 
Dashed and dashed-dotted lines are the results of AL when $m_*/m = t/t_{\rm eff}$, see Eq.~(\ref{teff}).
Error bars are shown but are typically smaller than symbol sizes.}
\label{mvsg2}
\end{figure}

\textit{Quantum Solution}.
Accurate solutions using XMC method can be obtained for any value of $g_2$.
In all cases considered, for $g_2 > g_{\rm AL}$ we observe perfect agreement with 
theoretical predictions for AL, see Figs.~\ref{dEvsg2}--\ref{mvsg2}. For realistic 
$\Omega /t = 0.1$ ratio, this regime corresponds to extremely large coupling $g_2 > 10^5$.
For $U =0$, $V =0$, the bipolaron state emerges at around $g_2 \approx 400$
with large but finite $R^2 \approx 40$, see Figs.~\ref{dEvsg2} and \ref{Rvsg2}.
Simulations at smaller $g_2$ yield energies $E$ that are slightly larger than the energy of two single polarons $2E_p$ and gradually decreasing with increasing the projection time $\tau$ in Eqs.~(\ref{G2}) and (\ref{Operator}). Concurrently, 
the measured mean-square radius is diverging with $\tau$. These are clear signatures
that the two-particle ground state is unbounded. 
Large radius bipolaron (and polaron) solitons form because they optimize the trade-off 
between the smaller positive interaction energy and larger localization energy 
for high electron density \cite{Kuklov,Kuklov2,Gogolin,X2soliton}. The value of $T_c$ in this regime
is small because of large $R_*^2$ (large-radius solitons remain relatively light when they form).
For larger values of $g_2$ solitons are getting smaller until they reach the AL. For $T_c$ this effect is more important than the slow increase of the effective mass, 
especially given that on the approach to the atomic limit the $m_*$ vs $g_2$ dependence 
exhibits a broad plateau (for a smaller adiabatic ratio it is even expected to go through a minimum \cite{X2soliton}), see Fig.~\ref{mvsg2}. 
The broad maximum in $T_c$ is reached for $R^2 \gtrsim 1$ because at stronger coupling the effective mass increase becomes the dominant effect in suppressing 
the transition temperature. 
At the maximum the $T_c/\Omega$ ratio is as large as $2$ 
and  by far exceeds all known best case scenarios for the linear EPI. However, one may not miss
the fact that this corresponds to very large values of $g_2$ when $\tilde{\Omega}_2$ is comparable to the bare electron bandwidth.

Since the bound state first emerges as a large-radius soliton state, 
it can easily tolerate strong Hubbard repulsion by developing local inter-particle correlations. As a result, for $(U/t,V/t)=(6,0)$ the bipolaron state emerges at slightly larger $g_2\approx 600$. The most important effect of onsite repulsion is significant shift
towards large $g_2$ for the onset of AL and the development of a broad plateau 
in both $R^2(g_2)$ and $m_*(g_2)/m$, see Figs.~\ref{Rvsg2} and (\ref{mvsg2}). The maximum of $T_c$ is found at the end of this plateau (see in Fig.~\ref{Tcvsg2}).

On the one hand, while remaining stable against Hubbard $U$ large bipolarons most likely form an attractive Bose gas, i.e. the system is unstable against phase separation. On the other hand, compact bipolarons (certainly in the AL) repel each other because of the Pauli principle \cite{PhysRevLett.132.226001}. The phase separation scenario is eliminated by adding Coulomb repulsion, but our simulations show that even relatively weak Coulomb potential with $V/t=0.5$ shifts the threshold for bipolaron formation by orders of magnitude! Now bipolarons 
first emerge at $g_2\approx 2 \times 10^4$  as small size, $(R/a)^2 \sim 3$, states.
However, their properties and subsequent evolution towards AL 
are nearly independent of $V$, see Figs.~\ref{Tcvsg2}-\ref{mvsg2}. This outcome ensures
that in the region of $T_c$ maximum bipolarons form a stable superconducting state. 
  

\textit{Conclusions}.
We present numerically exact results for bipolarons from quadratic EPI with and without the electron-electron interaction in the adiabatic regime with $\Omega /t =0.1$. Our data cover the entire strong coupling regime from the threshold of forming a shallow large-radius soliton state to 
its evolution towards compact bipolarons, and, ultimately, to the atomic limit when electrons 
are tightly bound on a single site. The effective masses of bipolarons remain relatively light
and exhibit extremely weak dependence on the EPI until the atomic limit is reached. This leads to
a maximum of $T_c$ estimated from Eq.~(\ref{Tc3DR}) on approach to the atomic limit when 
$(R/a)^2 \gtrsim 1$. At the maximum the $T_c/\Omega$ ratio is much larger than what is possible to achieve in all known models with the linear EPI, but the quadratic coupling must be 
such that the locally renormalized phonon frequency be comparable to the bare electron bandwidth. Realistically, this situation may describe quadratic EPI to a soft (or near critical) mode, $\Omega \to 0$  at $g_2 \Omega^2 $=const.
This outcome also suggests that high-$T_c$ mechanism is easier to get for larger $\Omega /t$ ratios.  
The XMC method used to solve the problem, can be generalized to models with arbitrary density-displacement coupling, arbitrary sign-positive displacement dependent hopping $t(x)$, and dispersive phonons \cite{Xrep} to investigate whether even higher values of $T_c$ are feasible within the bipolaron mechanism of superconductivity.

The essence of the binding mechanism discussed here is minimization of the phonon zero-point energy. It turns out that thermal fluctuations \textit{increase}
the free energy balance in favor of bipolarons and the thermal contribution 
becomes dominant at $T > \Omega\sqrt{g_2} /\ln(g_2)$, i.e. at 
temperature much lower than the ground-state binding energy. This 
motivates  further studies of finite-$T$ effects; in particular, the extent of the 
preformed pairs regime.

AK, NP and BS acknowledge support from the National Science Foundation under Grants 
DMR-2335905 and DMR-2335904, 
ZZ acknowledges support from the National Science Foundation under Grant DMR-2032077 and from Simons collaboration on new frontiers in superconductivity.
This work was performed in part at Aspen Center for Physics, which is supported by National Science Foundation grant PHY-2210452.

\bibliography{reference}
\end{document}